\documentclass[
aps,
pre,
twocolumn,
amsmath,
amssymb,
showpacs,
floatfix,
letterpaper
]{revtex4-1}
\usepackage[english]{babel}
\usepackage{graphicx}
\usepackage{bm}
\usepackage{amssymb,amsmath}
\usepackage{dcolumn}
\usepackage{wasysym}

\begin{document}

\title{Distribution of local relaxation events in an aging 3D glass: spatio-temporal correlation and dynamical heterogeneity}

\author{Anton \surname{Smessaert}}
\author{J\"{o}rg \surname{Rottler}}
\affiliation{Department of Physics and Astronomy, The University of British Columbia, 6224 Agricultural Road, Vancouver, British Columbia, Canada, V6T 1Z1}

\begin{abstract}
We investigate the spatio-temporal distribution of microscopic relaxation events, defined through particle \emph{hops}, in a model polymer glass using molecular dynamics simulations. We introduce an efficient algorithm to directly identify hops during the simulation, which allows the creation of a map of relaxation events for the whole system. Based on this map, we present density-density correlations between hops and directly extract correlation scales. These scales define collaboratively rearranging groups of particles and their size distributions are presented as a function of temperature and age. Dynamical heterogeneity is spatially resolved as the aggregation of hops into clusters, and we analyze their volume distribution and growth during aging. A direct comparison with the four-point dynamical susceptibility $\chi_4$ reveals the formation of a single dominating cluster prior to the $\chi_4$ peak, which indicates maximally correlated dynamics. An analysis of the fractal dimension of the hop clusters finds slightly non-compact shapes in excellent agreement with independent estimates from four-point correlations.
\end{abstract}

\date{\today}
\maketitle

\section{Introduction}

The connection between large scale structural relaxation and microscopic dynamics is a central component towards understanding the properties of amorphous solids. When such glassy materials are formed through a rapid temperature quench from the liquid phase, their viscosity increases by many orders of magnitude over a small temperature range around the glass transition temperature $T_g$~\cite{debenedetti_supercooled_2001}. Below this temperature, the dynamics have sufficiently slowed to prevent equilibration of the glass on experimental time scales. Non-equilibrium relaxation processes continue to evolve the system towards energetically favorable configurations, and material properties like density or mechanical response change with time, i.e. are history dependent. This so called "physical aging" has been studied extensively in many structural glasses. Beginning with the seminal studies of Struik 35 years ago on polymers~\cite{struik_physical_1978}, experiments consistently find that during aging, bulk quantities such as density and enthalpy increase logarithmically with the time elapsed since the glass was formed, while the principal structural relaxation time increases as a power law with age. This $\alpha$-relaxation time, which can be directly observed from the decay of intermediate scattering functions~\cite{kob_aging_1997} in computer simulations, ties the molecular scale dynamics to the non-equilibrium evolution and can be used as an "internal clock" of the glass~\cite{smessaert_recovery_2012}. The universal features of aging are an essential ingredient for the development of predictive models of the mechanics of glasses~\cite{chen_molecular_2007,chen_theory_2011} and are intimately tied to the glass transition itself.

Many earlier studies on either side of the glass transition focused on spatially averaged quantities, yet conclusive insights into the underlying physics have proven to be elusive. Especially the absence of diverging response functions close to the transition has prevented the identification of a static correlation length associated with collective processes.  The discovery of dynamical heterogeneity (DH), both experimentally~\cite{ediger_spatially_2000,berthier_direct_2005,cipelletti_slow_2005} and computationally~\cite{hurley_kinetic_1995,kob_dynamical_1997,lacevic_spatially_2003,chaudhuri_universal_2007} redirected the efforts to the study of local quantities. The notion of DH refers to the fact that unlike in a liquid, glasses exhibit regions of high and low molecular mobilities. First insights were gained by computational studies that monitored a subset of "fast" particles, which revealed a heterogeneous distribution and string-like cooperative motion~\cite{kob_dynamical_1997,donati_stringlike_1998}. To study the correlation between the dynamics at different locations in the glass, four-point correlation functions were introduced~\cite{bennemann_growing_1999}, which allowed the measurement of a "dynamical" susceptibility $\chi_4$~\cite{lacevic_spatially_2003,berthier_direct_2005}. A key result was the observation that $\chi_4$ is increasing in magnitude as temperature is approaching the glass transition from above. This increasing DH is now widely believed to provide a missing access point to the processes underlying the transition and provides direct evidence for a growing "dynamical" correlation length (see Refs.~\cite{berthier_theoretical_2011,berthier_dynamical_2011} for current reviews).

The present computational study aims at observing the correlated dynamics and DH of a glass in the aging regime, where only a few studies exist~\cite{parsaeian_growth_2008}. We establish a spatially resolved measure of DH that complements the standard bulk averaged dynamical susceptibility $\chi_4$ and analyze the non-equilibrium evolution (aging) from both perspectives. To identify heterogeneities, many earlier studies reduced the analysis to a subset of (im)mobile particles. However, this makes drawing of conclusions for the whole system challenging. The microscopic dynamics in a glass allow an alternative route to a reduced picture. The particle motion is dominated by periods of localized vibrational motion in the shell of surrounding particles, that is separated by rapid changes in the trajectory when a particle escapes the shell and \emph{hops} to a new position. Vollmayr-Lee et al.~\cite{vollmayr-lee_single_2004,vollmayr-lee_heterogeneities_2005,vollmayr-lee_self-organized_2006,vollmayr-lee_microscopic_2013} were amongst the first to exploit this essential feature of glassy dynamics to measure the effective motion in a glass. They confirmed string-like cooperative motion and found indications for an age- and temperature-independent distribution of "persistence times," i.e., times that particles stay inactive between hops. A closer investigation of these persistence times by Warren and Rottler in quiescent polymer- and binary Lennard-Jones glasses~\cite{warren_atomistic_2009} showed directly how aging arises from a broad distribution of relaxation times and can be quantitatively described with continuous time random walks. Subsequent studies of mechanically deformed glasses~\cite{warren_microscopic_2010,warren_deformation-induced_2010} revealed a narrowing of the persistence time distribution which in turn was linked to accelerated dynamics (mechanical rejuvenation). These studies, however, could only deduce statistical information, since only a randomly chosen subset of particles were studied. Candelier et al. succeeded in monitoring all hopping particles in experimental setups of agitated granular media~\cite{candelier_building_2009,candelier_dynamical_2010} and a simulation of a supercooled fluid~\cite{candelier_spatiotemporal_2010} in two dimensions, and they used this information to directly study the correlation between hops. They found a hierarchy of aggregations into avalanches that are directly linked to DH, and related their size and position to soft regions identified via local Debye-Waller factors. Yet their detection method is impractical for large numbers of particles, limiting it's applicability mostly to two-dimensional (2D) systems.

We present an improved algorithm that allows the measurement of hops for all particles in large three-dimensional (3D) glasses at finite temperature. We apply this algorithm in molecular dynamics simulations of a standard bead-spring polymer model in the aging regime, well below the glass transition temperature. A polymer model is chosen since aging effects are best explored for this type of glass, yet our results should be indicative for other structural glasses such as colloidal or metallic glasses as well. The employed polymer chain length is well below the entanglement length, and the hop process is dominated by the crowded surroundings rather than the presence of polymer topology~\cite{aichele_polymer-specific_2003}. The detection of hops throughout the system gives us a map of local relaxation events, and in the first part of this work we analyze hop statistics and compare results to earlier studies. Furthermore, we directly calculate the density-density correlation between hops, which allows us to infer correlation ranges, and these are used to quantitatively analyze collaborative rearrangements of particles. In the second part of this study, we focus on DH in the aging regime. This regime has only recently been explored using three- and four-point correlators~\cite{parsaeian_growth_2008,brun_evidence_2012}, finding indications for an increase of dynamical correlation with age. On the basis of our hop detection, we calculate the four-point dynamical susceptibility $\chi_{4}$ as the standard measure for correlated dynamics and directly compare it with the aggregation of hops into clusters, which is the manifestation of DH in the hop picture. With this spatial resolution of DH we calculate the full volume distribution of hop clusters and give a measure of their compactness.

\section{Methods}\label{methods}
We simulate a polymer glass using the molecular dynamics technique and the well-known finitely extensible nonlinear
elastic (FENE) bead-spring model~\cite{kremer_dynamics_1990}. It consists of a 6-12 Lennard-Jones (LJ) potential and a stiff non-linear spring-like interaction between bonded particles that prevents chain crossings. This model is used extensively in the study of glassy systems and exhibits key features like a power-law age dependence of the relaxation time and logarithmic aging of structural properties~\cite{smessaert_recovery_2012}. All results are reported in the usual LJ units, based on energy well depth $\epsilon$, particle diameter $\sigma$, and mass $m$, as well as the characteristic time scale $\tau_{LJ}=\left( m \sigma^2/\epsilon\right)^{1/2}$. We simulate 1000 chains, consisting of 50 particles each, in a cubic box with periodic boundaries and at a density that is chosen to yield a system close to zero pressure at the end of the quench. The glasses are generated with a constant volume quench from an equilibrated melt at $T=1.2\epsilon/k_B$ (units of temperature omitted in the following) at the constant rate of $6.7\cdot10^{-4}\tau_{LJ}^{-1}$ to the final temperatures $T=0.2,0.25,0.3$, which are below the glass temperature $T_g\simeq 0.35$~\cite{rottler_shear_2003}. In a final step the pressure is ramped to zero in $37.5\tau_{LJ}$. After this, we study the evolution of the quiescent glass at $T=const$ and fixed pressure $P=0$ during $10^8$ time steps. The simulations are performed with LAMMPS~\cite{plimpton_fast_1995}, using a Nos\'e-Hoover thermostat and barostat, and a time step of $\Delta t = 0.0075\tau_{LJ}$. To increase performance the potentials are truncated at $r_c=1.5\sigma$ and shifted for continuity. To increase the accuracy of our analysis, especially those in the second results section, we performed 11 independent simulation runs for temperatures $T=0.25,\,0.3$ and report averaged results. Error bars indicate the standard error assuming a Gaussian distribution. Results for $T=0.2$ are generated from a single run, but are otherwise calculated identically. Age-dependent results, where not otherwise specified, were calculated in an observation time window much less than the age itself $[t_{age},1.1\cdot t_{age}]$.

\begin{figure}[tb]
\includegraphics{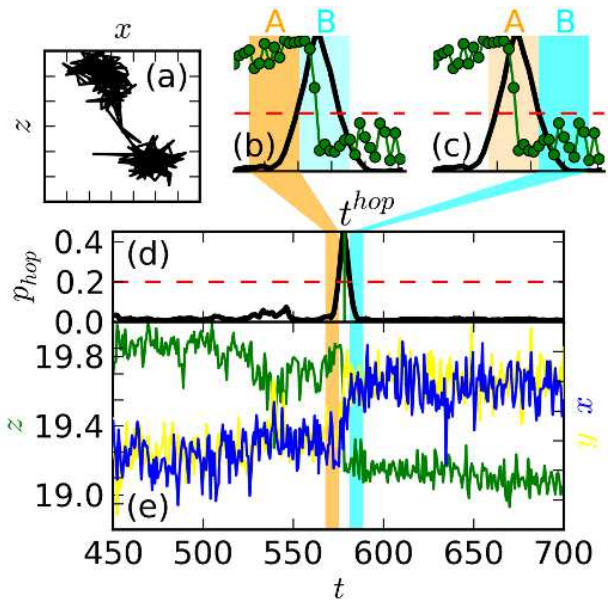}
\caption{\label{fig:trajectory}(Color online) Sample trajectory of a particle and hop identifier function $p_{hop}$. (a) shows the localization into two cages and the 'hop' is marked by rapid changes in the trajectory (e). Corresponding to this, $p_{hop}$ in (d) is sharply peaked at the transition and the maximum defines the hop time $t_{hop}$. Plots (b) and (c) are overlays of $p_{hop}$ and the trajectory just before and after the hop (z comp. only for better visibility). The colored fields (A and B) in both plots indicate the evaluation window for $p_{hop}$ [see Eq.~\eqref{eq:hopprobability}]. Initial and final position of the particle is calculated from time averages of the trajectory sections that are highlighted as zoom, i.e., (b) initial position (orange) and (c) final position (cyan).}
\end{figure}
In this study we focus on the cage-escapes of the particles and we record these \emph{hops} by employing an adaptation of the algorithm proposed by Candelier et al.~\cite{candelier_building_2009}. During the simulation we store the most current part of the trajectory of length $t_{eval}=15\tau_{LJ}$ (see below for details) for each particle. This trajectory is separated into two sections of equal size \textit{A} $(t-t_{eval},t-t_{eval}/2]$ and \textit{B} $(t-t_{eval}/2,t]$, and a \textit{hop identifier function} $p_{hop}(t-t_{eval}/2)$ is calculated every 100 time steps

\begin{equation}\label{eq:hopprobability}
p_{hop}(t-\frac{t_{eval}}{2}) = \sqrt{\left\langle \left(\mathbf{r}^A - \bar{\mathbf{r}}^B\right)^2\right\rangle_A\cdot \left\langle \left(\mathbf{r}^B - \bar{\mathbf{r}}^A\right)^2\right\rangle_B}<p_{th}\; .
\end{equation}

This function is a measure of the averaged distance between the mean position in section \textit{A}, $\bar{\mathbf{r}}^A$, and all trajectory points in section \textit{B}, $\mathbf{r}^B$, and vice versa. It  is large when the trajectory changes rapidly at around $t-t_{eval}/2$. We define that a hop is detected when $p_{hop}$ exceeds a temperature dependent threshold, and in this study we use $p_{th}=0.21(T=0.3),\;0.18(0.25),\;0.15(0.2)$ (for details see below). Our adaptation to the original algorithm is to use a running evaluation time window $t_{eval}$ instead of a recursive scheme on the whole trajectory and this allows us to detect hops \emph{on-the-fly} during the simulation run. In Figs.~\ref{fig:trajectory}(a) and (e) we show a sample trajectory of a particle together with the calculated $p_{hop}$, Fig.~\ref{fig:trajectory}(d). A hop is clearly identified and the algorithm records particle id, hop time $t_{hop}$ and position of the particle before and after the hop $\mathbf{r}_{init}$ and $\mathbf{r}_{final}$. The hop time is defined at the maximum of $p_{hop}$, and the locations are calculated from averages as the threshold is crossed: before the peak $\mathbf{r}_{init}=\langle \mathbf{r} \rangle_A$ [see Fig.~\ref{fig:trajectory}(b)] and after the peak $\mathbf{r}_{final}=\langle \mathbf{r} \rangle_B$ [see Fig.~\ref{fig:trajectory}(c)]. Using this algorithm on each particle via a parallel implementation in LAMMPS~\cite{plimpton_fast_1995}, we are able to monitor all hops in the system for the full duration of the simulation. The following paragraphs provide details on the detection algorithm and explain the parameter-choices. Readers that are mainly interested in the results are encouraged to proceed to the next section.

\begin{figure}[tb]
\includegraphics{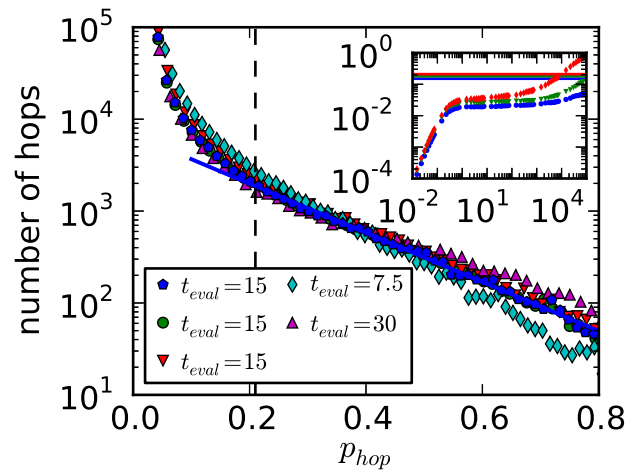}
\caption{\label{fig:hopHistogram}(Color online) Histograms of hop identifier $p_{hop}$ peak heights calculated at glass temperature $T=0.3$ and time window parameters (order of legend):$(N_{hist},N_{obs})=(20,100),\,(40,50),\,(10,200),\,(10,100),\,(40,100)$. The data for the three parameter sets with $t_{eval}=15\tau_{LJ}$ are nearly identical (blue $\pentagon$ overlay other markers). The solid line is an exponential fit of the distribution tail and the dashed line indicates the resulting parameter choice $p_{th}$. The inset shows the mean square displacement for temperatures $T=0.2$(blue $\bigcirc$), $T=0.25$(green $\bigtriangledown$), and $T=0.3$(red $\Diamond$). The horizontal lines indicate $p_{th}$ at the corresponding temperature.}
\end{figure}
The algorithm for the detection of hop events is centered around a threshold criterion for the distance between two adjacent parts of the trajectory, see Eq.~\eqref{eq:hopprobability}. A cage-escape happens whenever this distance is bigger than a cage size $\sigma_{c}$, and Candelier et al.~\cite{candelier_building_2009,candelier_dynamical_2010} argued that the mean cage size is known from the height of the plateau in the mean square displacement of the particles. Therefore, a threshold $p_{th} \gtrsim \sigma_{c}^{2}$ was proposed and successfully used in studies of agitated granular media~\cite{candelier_building_2009,candelier_dynamical_2010}, a supercooled liquid in 2D~\cite{candelier_spatiotemporal_2010} and a cyclically sheared glass in 3D~\cite{priezjev_heterogeneous_2013}. We follow this argumentation and furthermore attempt to reduce the ambiguity in the choice of $p_{th}$ by measuring a histogram of peak-heights in the hop identifier $p_{hop}$. Assuming that a particle has only two well defined modes of movement: vibration around a fixed mean position (caged) and instant jump by a distance $>\sigma_{c}$ (hop), then the hop detection would only pick up peaks of height $>\sigma_{c}^{2}$. In Fig.~\ref{fig:hopHistogram} the blue($\pentagon$) data (standard time window parameters, see below) are the measured histogram. We find a distribution of peak heights that is largest at small heights, with a long exponential tail towards larger peaks. The small peaks are due to shifts of the mean particle position within the cages. This is expected, because each member of the surrounding shell that form the cage is not fixed in space, but moves within a small area (its own cage). Therefore, the cages are not static, but change slightly over time. The measured distribution shows a relatively sharp transition to a slow exponential tail, which indicates the existence of an additional source of larger peaks. These peaks are caused by the escape of particles from their cages and we therefore use the point of transition as our hop threshold. Based on the data in Fig.~\ref{fig:hopHistogram} we use $p_{th}=0.21$ at $T=0.3$ (dashed line), and equivalent arguments provide thresholds $p_{th}=0.18$ and $p_{th}=0.15$ for the temperatures $T=0.25$ and $T=0.2$. The inset shows the mean square displacement for all three temperatures and the horizontal lines indicate the thresholds used in this study.

The time window is introduced into the algorithm similar to a running average scheme: During the simulation we store $N_{hist}$ trajectory points for each particle. Every $N_{obs}$ time steps the oldest point is replaced and $p_{hop}$ recalculated. We choose $N_{obs}=100$, which ensures a reevaluation of the hop detection every mean collision time ($N_{obs}\cdot\Delta t=0.75\tau_{LJ}$ see the inset in Fig.~\ref{fig:hopHistogram}), making this our maximal temporal resolution for the hop time. The length of the time window $t_{eval}=N_{hist}\cdot N_{obs}\cdot\Delta t$ is the main influence on the detection algorithm. It sets the maximal resolution of two consecutive hops ($t_{eval}/2$) and also acts as an upper limit for the duration of a hop. To maximize the resolution but also ensure that meaningful averages are possible [see Eq.~\eqref{eq:hopprobability}], we choose $N_{hist}=20$ and therefore $t_{eval}=15\tau_{LJ}$. In fig.~\ref{fig:hopHistogram} we illustrate the impact of this parameter. Note that the histogram for our chosen parameter values (blue $\pentagon$) coincides with two other parameter sets with equal $t_{eval}$ (data are nearly identical and therefore difficult to see), indicating that only the window size has a direct impact on the hop detection. Consequently, a smaller time window yields the detection of more peaks with small height, and a larger $t_{eval}$ results in fewer small peaks, but also in a lower resolution.

\section{Results}
\begin{figure}[tb]
\includegraphics{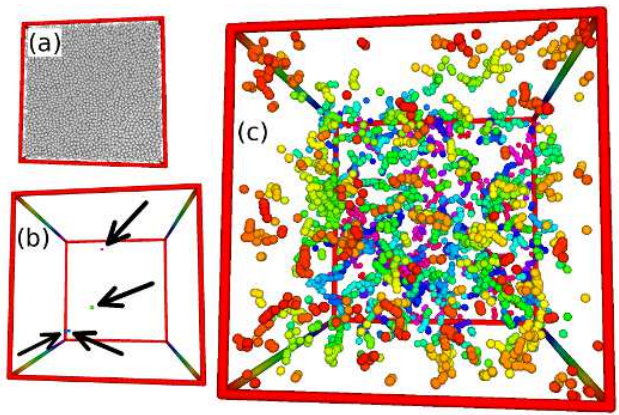}
\caption{\label{fig:hopsnapshot}(Color online) Snapshots of a single configuration showing (a) all particles and (b) only those particles that are in the middle of a hop. There are only four hops at that time step and their position is highlighted by arrows. (c) shows all hops that are detected in a time window of $3000\tau_{LJ}$. The configurations are taken from a glass at $T=0.2$ and age $20000\tau_{LJ}$.}
\end{figure}
The detection algorithm developed in this study allows us to record a list of all cage-escapes or \emph{hops} that happen during a simulation run. Each hop is characterized by a time, the initial and final position of the particle (see previous section for definitions) and its id. In Fig.~\ref{fig:hopsnapshot}(a) we show a snapshot of the whole system followed in Fig.~\ref{fig:hopsnapshot}(b) by a reduced picture where only particles are shown that hop at that time step. The comparison highlights the sparseness of the effective dynamics, showing a reduction from 50000 to four particles. By merging all hops that were detected in a time window of $3000 \tau_{LJ}$ we directly reveal in Fig.~\ref{fig:hopsnapshot}(c) the heterogeneous distribution of the hops and their grouping into clusters. In other words, we directly show the dynamical heterogeneity in the glass.

\subsection{Statistical properties of hops}
\begin{figure}[tb]
\includegraphics{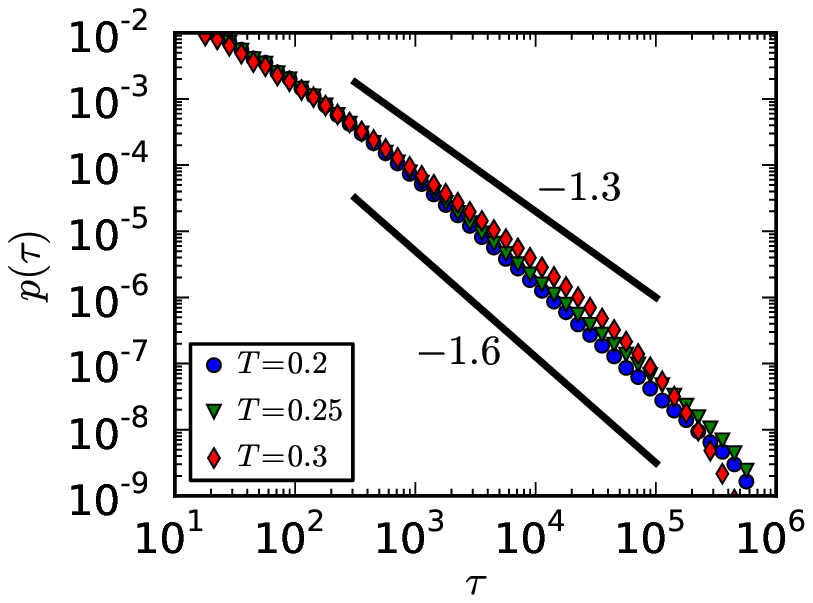}
\includegraphics{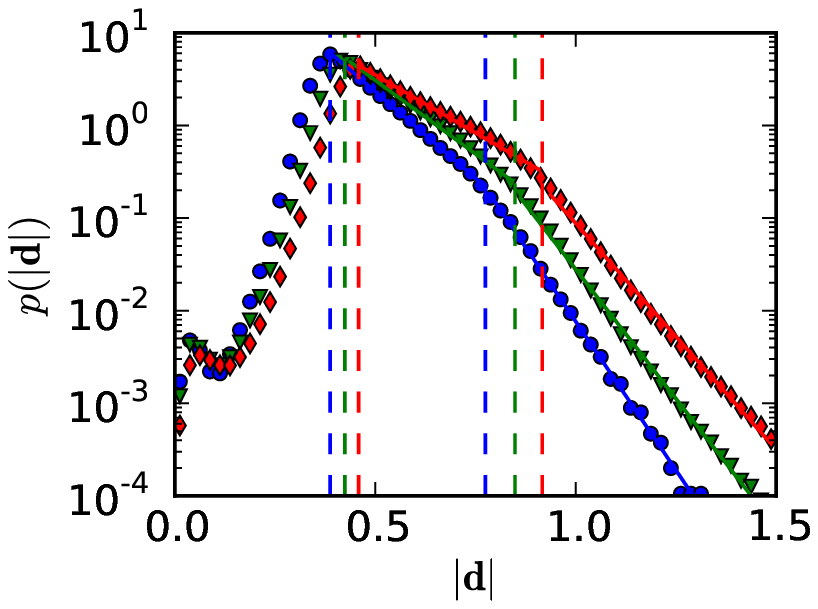}
\caption{\label{fig:persisencetime}(Color online) Top: Distribution of persistence time $\tau$ of cages at three temperatures. The solid lines indicate power laws. Bottom: Distribution of hop distances $|d|$, i.e., the distance between old and new cage. Exponential fits are indicated by the solid lines and the vertical dashed lines indicate $\sqrt{p_{th}}$ and $2\sqrt{p_{th}}$ for the respective temperatures; see legend in top panel. Error bars in both plots are smaller than the markers and are omitted for visibility.}
\end{figure}
A key feature of glasses is their non-equilibrium nature, giving rise to a continuously slowing structural relaxation with increasing age. The mean squared displacement exhibits a plateau that indicates the trapping of particles into local cages (see also inset in Fig.~\ref{fig:hopHistogram}), and the length of the plateau increases with age. This suggests an increasing persistence time in the cage, i.e. the time between two consecutive hops of the same particle. This aging is caused by a broad distribution of persistence times and previous studies have found a power law $p(\tau)\propto \tau^\mu$ with exponent $-1 > \mu > -2$ in simulations of polymer glasses and binary mixtures~\cite{warren_atomistic_2009}. In a very recent study this broad distribution was also found in simulations of a strong glass former~\cite{vollmayr-lee_microscopic_2013}, although with exponent $-0.3 \geq \mu \geq -1$. In Fig.~\ref{fig:persisencetime} we show the persistence time distribution for quiescent glasses at three temperatures $T=0.2,0.25,0.3$. We find a power-law behavior with $\mu\simeq -1.5$ insensitive to the glass temperature, which agrees well with the previously found value for polymer glasses of $-1.23$~\cite{warren_atomistic_2009}. The slightly smaller value is probably due to the increased sensitivity of our algorithm, which will increase the likelihood of shorter persistence times. The shortened tail observed for $T=0.3$ indicates that persistence times in this system are sampled from a finite distribution and hence the polymer will equilibrate at long times (see also Ref.~\cite{vollmayr-lee_microscopic_2013}). However, since the turnover happens at times of the order of the total simulation time, we are still well within the aging regime.

During a hop, the particle moves from one metastable local configuration to another, i.e., from one cage to the next. Our algorithm estimates both locations $\mathbf{r}_{init},\,\mathbf{r}_{final}$ and we can therefore calculate the hop distance
\begin{equation}\label{eq:hopdistance}
|\mathbf{d}|=|\mathbf{r}_{final}-\mathbf{r}_{init}|\; .
\end{equation}
In the bottom panel of Fig.~\ref{fig:persisencetime} the distribution of the hop distance is shown for three temperatures. For each glass, the main peak is located at $\sqrt{p_{th}}$ (the first vertical dashed line), which is the minimal distance that the detection algorithm sets for the separation of two ideal cages. We find an exponential decay following the peak, with a transition to a faster exponential decay at around $2\sqrt{p_{th}}$. The transition is expected, because at distances $>2\sqrt{p_{th}}$ it is possible for the detection algorithm to separate the particle motion into two hops, if the particle briefly stabilizes at an intermediate distance. The form of the distribution is qualitatively unchanged for varying temperatures, suggesting that the hop process is unchanged inside the glass phase. Indeed, in simulations of a strong glass former~\cite{vollmayr-lee_microscopic_2013} a comparable distribution was found, indicating a similar role of the hop process.

\begin{figure}[tb]
\includegraphics{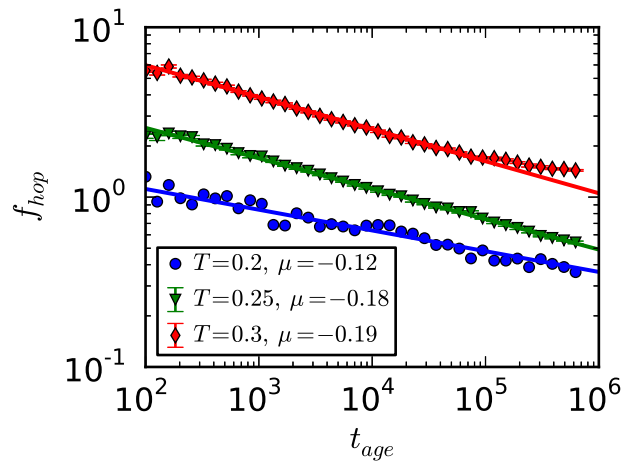}
\caption{\label{fig:hopfrequency}(Color online) The hop frequency $f_{hop}$ is the number of hops observed in the whole system per time $\tau_{LJ}$. Shown is the frequency over the duration of the simulation run in a log-log plot, with solid lines indicating power-law fits.}
\end{figure}
In Fig.~\ref{fig:hopfrequency} we show the hop frequency of the glass, i.e. the number of hops per time $\tau_{LJ}$. We observe about six times more hops at $T=0.3$ compared to $T=0.2$. Furthermore, we find that the frequency decreases with a weak power law with age. The aging is accelerated at higher temperatures, which indicates that the phase space is explored more quickly. Indeed, for the glass closest to the glass transition ($T=0.3$), we observe a flattening of the curve, i.e. the simulation reaches timescales close to the end of the aging regime.

\begin{figure}[tb]
\includegraphics{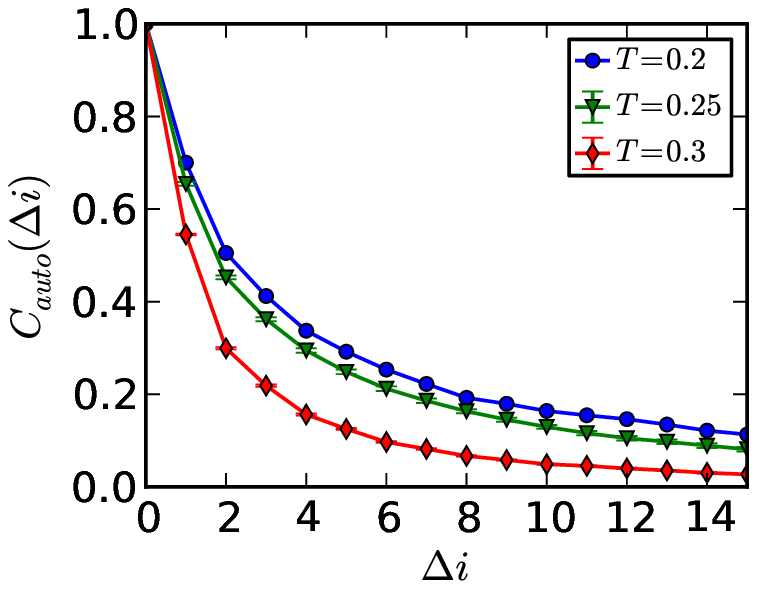}
\includegraphics{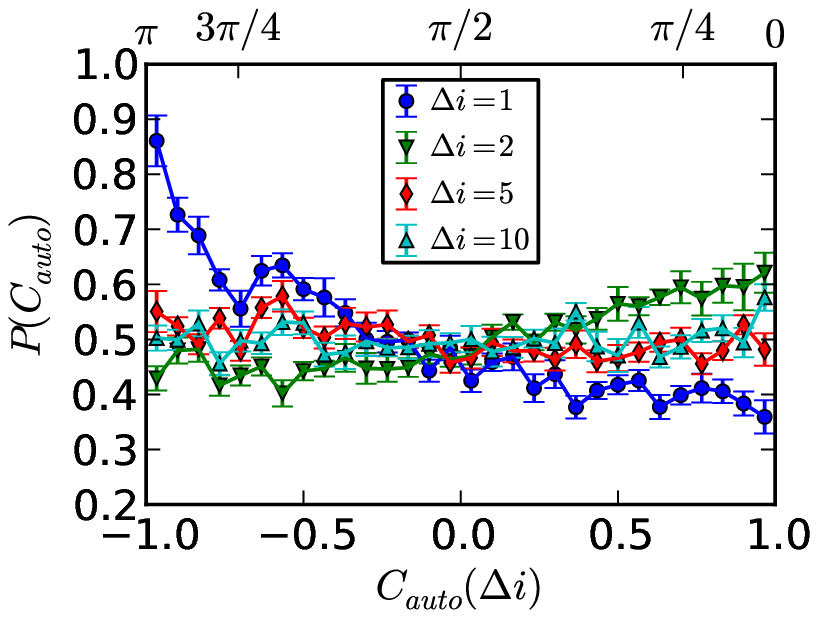}
\caption{\label{fig:autocorrelation}(Color online) Hop displacement autocorrelation for three glass temperatures (top) at age $t_{age}=10^{5}\tau_{LJ}$. The histogram (bottom) is calculated using normalized displacement vectors ($T=0.25$ and age as above) and reveals an anisotropy in the direction of consecutive hops. The lines are guides to the eye.}
\end{figure}
We can further characterize the hop process by calculating the autocorrelation
\begin{equation}
C_{auto}(\Delta i) = \frac{\langle \mathbf{d}_i\!\cdot\mathbf{d}_{i+\Delta i}\rangle}{\langle \mathbf{d}_{i}\cdot\mathbf{d}_{i}\rangle}
\end{equation}
of the displacement vector $\mathbf{d}$ [see Eq.~\eqref{eq:hopdistance}]. The average in the numerator is taken over all hop-pairs of a particle with a separation $\Delta i$ and the denominator is the variance of $\mathbf{d}$ with zero mean. Simple mean-field trap models~\cite{monthus_models_1996} assume a solely temperature driven escape from the cage, which yields independent hops, and a previous study indeed found a vanishing autocorrelation after about two hops~\cite{warren_atomistic_2009}. Although the earlier study used a different hop detection algorithm with lower sensitivity, our results shown in Fig.~\ref{fig:autocorrelation} principally agree with these findings. We observe a correlation that decreases below 0.2 after at most seven hops, with a more rapid decline at higher temperatures. The decay is slower than previously found, because our algorithm is able to separately pick up back-and-forth hops of a particle between the same two cages. We confirm this observation with the autocorrelation histogram in Fig.~\ref{fig:autocorrelation}, where the displacement vectors were normalized to unit length. This isolates the directional correlation of the hops and one can clearly see a pronounced anisotropy. For consecutive hops we find that angles close to $180^{\,\circ}$ are clearly favored, indicating that the particle is more likely to return to the location where it came from. Furthermore, there is an increased probability for a following third hop to be in the same direction as the first, indicating a back-and-forth between the two cages. The anisotropy is subdued with increasing separation of hops and vanishes at $\Delta i = 5$ for a glass at $T=0.25$.

\begin{figure*}[tb]
\includegraphics{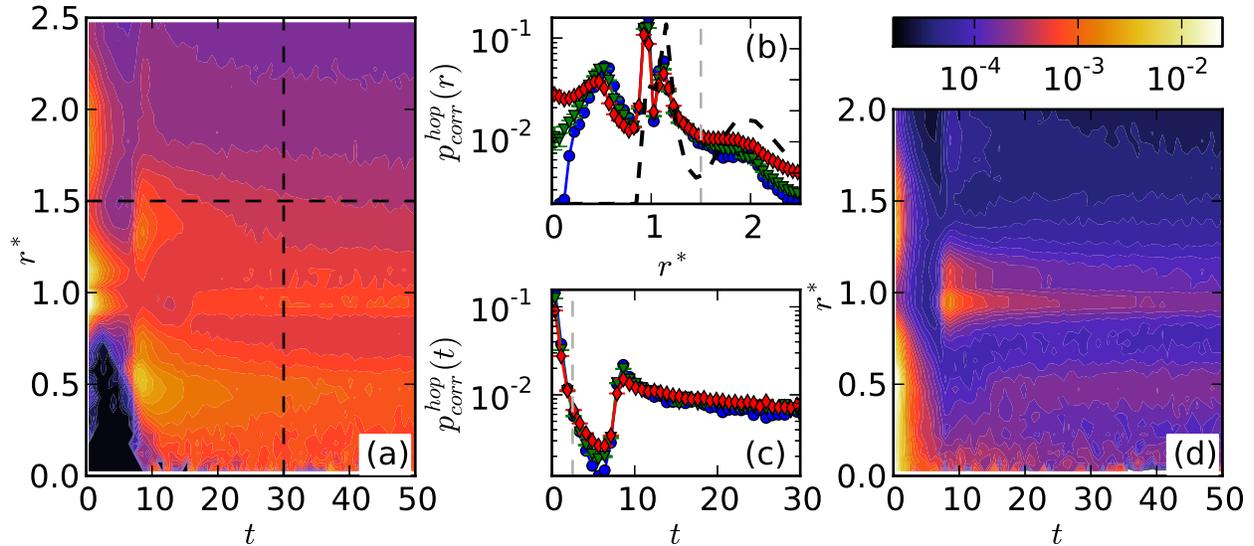}
\caption{\label{fig:hopcorrelation}(Color online) (a) Probability density surface for spatio-temporal separation of two different hopping particles based on Eq.~\eqref{eq:vanhove} for a glass at $T=0.25$ and $t_{age}=10^5\tau_{LJ}$. The color scale is logarithmic (scale at the top-right corner) and the dashed lines indicate integration limits used to calculate the one-dimensional probability functions (b,c). Center plots show the probability function of (b) separation and (c) time delay between hops for $T=0.2$(blue $\bigcirc$), $T=0.25$(green $\bigtriangledown$), and $T=0.3$(red $\Diamond$) at the same age. The gray vertical dashed lines in (b,c) illustrate the correlation ranges and the black dashed curve in (b) indicates the radial distribution function. (d) Probability density surface following Eq.~\eqref{eq:vanhove} for the same glass as the left panel, but with $r^*$ calculated from initial position (at the origin) to the final position of the second particle after the hop. The color scale is again logarithmic.}
\end{figure*}
Up to this point our results took advantage of the increased sensitivity and greater number of detected hops. However, since all particles are tracked, we are able to directly measure the spatio-temporal correlation of hops in a 3D glass. In Fig.~\ref{fig:hopcorrelation} we present surface plots that visualize probable temporal and spacial distances between hops. We calculate the density-density correlation of hops along the lines of the normalized distinct part of the van Hove function~\cite{hansen_theory_2006}
\begin{equation}\label{eq:vanhove}
G_d^{hop}(r,t) = \frac{1}{N^\prime}\left\langle \delta \left(r-|\mathbf{r}_{init}^{(i)}(0)-\mathbf{r}_{init}^{(j\neq i)}(t)|\right) \right\rangle \;.
\end{equation}
Here the average is taken over all $N^\prime$ hop pairs that involve two distinct particles, and using the initial positions, i.e. the initial cages. Figure~\ref{fig:hopcorrelation}(a) shows an example probability distribution. We find a dominating peak at times close to zero and at a distance $\simeq \sigma$, which is caused by hops in the surrounding shell of particles. The accumulation of near-simultaneous hops is a direct indicator of the cooperative nature of the hop process. The area $r\lesssim 0.7\sigma$ and $t\lesssim 7 \tau_{LJ}$ is empty, which is an effect of the excluded volume of the hopping particle at the origin. A secondary, at least an order of magnitude weaker peak is located around $\sigma/2$ and $10\tau_{LJ}$. It is caused by particles that hop after having entered the space that was vacated by the particle at the origin. Our data indicate that aging effects are minimal for the density-density correlation (not shown). To highlight the dependence on temperature we partially integrate Eq.~\eqref{eq:vanhove} from the origin to the dashed lines, which are chosen such that the main features are included. In Fig.~\ref{fig:hopcorrelation}(b) we show the spatial correlation $p_{corr}^{hop}(r)=\int_0^{30}\!\mathrm{d}t G_d^{hop}(r,t)$ for three temperatures. One can see that the main features are found at all temperatures and that an increased temperature weakens the sharpness of the peaks, which is due to the increased vibrational motion of the particles. Apart from the first peak at $\sigma/2$ (see above), we find peaks at positions that coincide with the static shell structure of the glass as indicated by the radial distribution function (black dashed curve). The splitting of the peak at around $\sigma$ is due to the different mean distance between particles that are neighbors in the same polymer backbone and particles that are not directly bonded to each other. The existence of the double peak shows that both pairs take part in cooperative rearrangements. Figure~\ref{fig:hopcorrelation}(c) shows the temporal correlation $p_{corr}^{hop}(t)=\int_0^{1.5}\!\mathrm{d}r G_d^{hop}(r,t)$. A sharp decay at small times is followed by a peak at around $9\tau_{LJ}$, which is due to the immediate re-hopping of particles; the back-and-forth hopping that is also discussed in connection with the hop autocorrelation (see above). The position of the peak, i.e. the secondary peak in the probability density surface is directly linked to the maximal resolution of two consecutive hops, which is $7.5\tau_{LJ}$ for the used parameters (see Sec.~\ref{methods}). We find only a very weak temperature dependence in the temporal correlation, suggesting that the fundamental mechanisms of cooperativity are the same over the temperature range studied here. Our data also do not show any clear indication for Poisson processes like those found for supercooled liquids and granular matter~\cite{candelier_building_2009,candelier_spatiotemporal_2010}. Based on the sharp drop following the main peak in the spatial probability distribution, we infer a correlation range of $r_{corr}=1.5\sigma$ [Fig.~\ref{fig:hopcorrelation}(b), vertical dashed line], i.e., the correlation does not extend beyond the nearest neighbor shell. From the temporal probability distribution we determine a correlation range of $t_{corr}=2.5\tau_{LJ}$ [Fig.~\ref{fig:hopcorrelation}(c), vertical dashed line], which is the time at which the initial peak has decayed to values below the close to constant region after the second peak. These ranges therefore restrict "correlated" hops to near-simultaneous hops of neighboring particles, as indicated by the primary peak in the density-density correlation.

Knowledge of the final position of hopping particles also allows us to explore the direction of correlated hops. The probability distribution surface in Fig.~\ref{fig:hopcorrelation}(d) shows a density-density correlation very similar to the one on the left. Again we use Eq.~\eqref{eq:vanhove}, but the distance is now calculated between initial cage at the origin and the \emph{final} position of other hopping particles $\mathbf{r}^{(j\neq i)}_{final}$. Therefore, high probability regions indicate where the particles end up after a correlated hop. By comparing the surfaces, we find that hops that started in the first shell (the primary peak in left plot) mostly end at $r\lesssim \sigma/2$, see the primary peak in the right plot. Indeed this peak extends all the way to $r\simeq 0$, indicating that it is possible for the cage at the origin to stay largely intact with a new particle taking the place of the last one. This suggests the string-like motion that was previously observed in a binary LJ mixture~\cite{donati_stringlike_1998}. We also find a secondary peak at a distance $r\sim 1.5\sigma$, which suggests that some hops from the first shell are directed away from the cage at the origin. Please note that the color scale in the plot is logarithmic, and that this second \textit{hop destination} is at least one order of magnitude less likely than the first one. The secondary feature in Fig.~\ref{fig:hopcorrelation}(d) at $\sim 10\tau_{LJ}$ is located around $\sigma$, confirming that the accumulation of hops at this time lag are due to back-hops of particles that return to the first shell after having hoped into the vacated volume closer to the center of the cage (origin).

\begin{figure}[tb]
\includegraphics{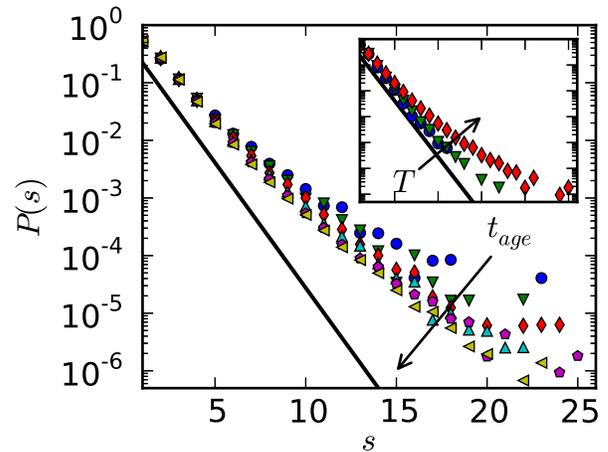}
\caption{\label{fig:colSize}(Color online) Size distribution of cooperatively rearranging particles. The main panel shows distributions at six ages for a glass at temperature $T=0.3$; see legend in Fig.~\ref{fig:chi4}. The inset shows the size distribution of three glasses at age $t_{age}=10^5\tau_{LJ}$ and temperatures $T=0.2$ (blue $\bigcirc$), $0.25$ (green $\bigtriangledown$), $0.3$ (red $\Diamond$). Both plots have the same axes ranges, and the solid black lines indicate $P(s)\propto \exp{(-s)}$.}
\end{figure}
The correlation ranges $r_{corr}=1.5\sigma$ and $t_{corr}=2.5\tau_{LJ}$ allow us to identify "cooperatively rearranging" groups of particles and we perform a cluster analysis to measure their size. Two particles are in the same cluster if they are closer in space and time than $r_{corr}$ and $t_{corr}$, which is in close analogy to Candelier et al.'s study~\cite{candelier_building_2009}. In Fig.~\ref{fig:colSize} we show the measured cluster size distributions, which exhibit initial exponential decays with stretched tails that approach a power law. In the main panel we show results for a single temperature and varying age, and one can see that as the age increases the distribution flattens and becomes more exponential-like. In an older glass the hop-activity is reduced (see hop frequency in Fig.~\ref{fig:hopfrequency}), and therefore the constant clustering time ($t_{corr}$) used here results in a lower likelihood of finding larger clusters with growing age. Varying temperature at the same age has a similar effect: as temperature increases, we observe a broadening of the distribution away from exponential and towards a power-law form.

\subsection{Dynamical heterogeneity and clustering of hops}
In previous studies the heterogeneous dynamics in glasses and supercooled liquids were mainly probed with four-point correlation functions. A standard approach is to measure the number of particles that remain approximately stationary as a function of time using overlap functions~\cite{lacevic_spatially_2003}. The variance of this quantity over a multitude of independent simulations is the four-point dynamical susceptibility $\chi_4$, which exhibits a peak at the time of maximal dynamical correlation in the system. The peak height is connected to the number of particles with correlated dynamics, and the increase of that height when approaching the glass transition signifies growing dynamical length scales. With our knowledge of the location and time of all hops in the system, we can provide a new perspective on the correlated dynamics. Specifically, we are able to spatially resolve the clustering of hops, directly revealing the heterogeneous dynamics, and study the cluster distribution as a complimentary perspective to $\chi_4$. Here we focus specifically on the aging regime, for which few studies exist. The results shown below are calculated from a glass at $T=0.3$, yet an equivalent analysis for $T=0.25$ confirms our findings further inside the aging regime.

\begin{figure}[tb]
\includegraphics{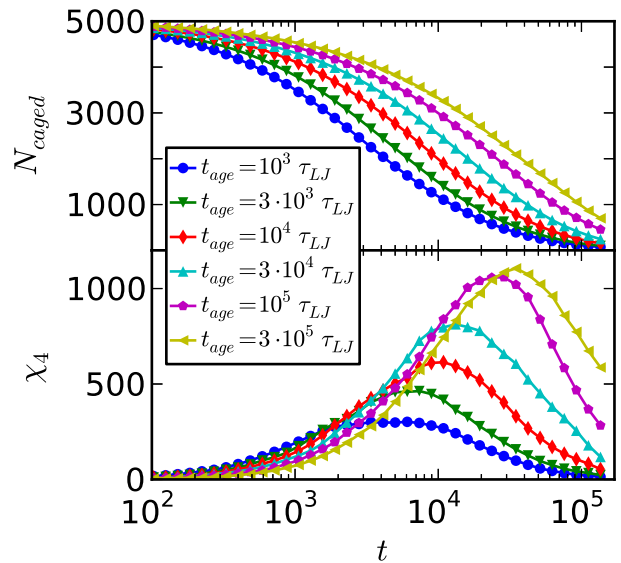}
\caption{\label{fig:chi4}(Color online) The top panel shows the number of caged, i.e. not yet hopped, particles $N_{caged}$ averaged over independent simulations as a function of time for six glass ages. The four-point susceptibility $\chi_4$ shown in the bottom panel is calculated from the variance of $N_{caged}$, Eq.~\eqref{eq:ncaged}.}
\end{figure}
In a first step, we calculate the number of particles that have \emph{not} hopped directly from the hop data
\begin{equation}\label{eq:ncaged}
	N_{caged}(t,t_{age}) = \sum_{i}^{N} b_{i}(t,t_{age})
\end{equation}
where $b_{i}(t,t_{age})=0$ if particle $i$ has hopped in the time window $[t_{age},t_{age}+t]$ and $b_{i}(t,t_{age})=1$ otherwise. In the upper panel of Fig.~\ref{fig:chi4}, we show results for six ages. We employed simulations of $N=5000$ particles that are otherwise equivalent to our usual simulations with $N=5\cdot10^{4}$ particles. We had to downscale our system, because converged measurements of $\chi_{4}$ required 300 independent runs. As mentioned above, the four-point dynamical susceptibility is proportional to the variance of $N_{caged}$~\cite{lacevic_spatially_2003}
\begin{equation}\label{eq:chi4}
	\chi_{4}(t,t_{age}) = \frac{\beta V}{N^{2}}\left( \langle N_{caged}^{2}\rangle - \langle N_{caged}\rangle^{2} \right)
\end{equation}
and the bottom panel of Fig.~\ref{fig:chi4} shows results for the same six ages. In an earlier study, Parsaeian and Castillo~\cite{parsaeian_growth_2008} investigated four-point correlations in the aging regime of a binary LJ glass, and we observe the same main features in Fig.~\ref{fig:chi4}: a shift of the peak towards larger times with increasing age and an increase in height, indicating a larger volume of correlated dynamics.

\begin{figure*}[tb]
\includegraphics{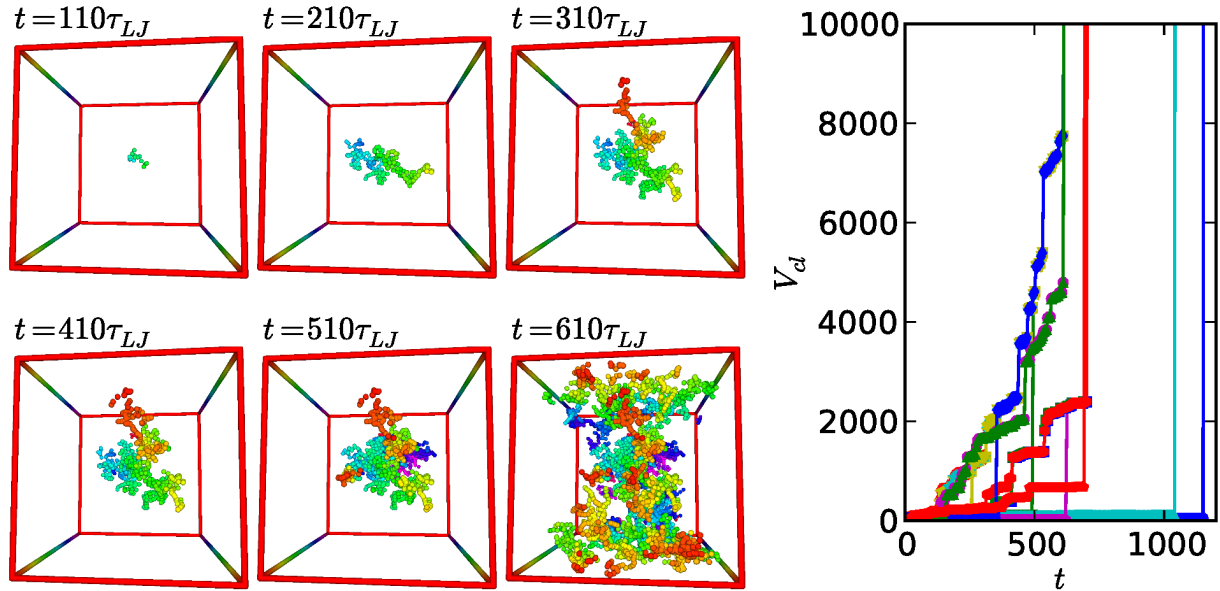}
\caption{\label{fig:growth}(Color online) Snapshots of the growth of a single cluster over time. The particles are visualized at their initial positions (before the hop) and the coloring indicates depth. The plot on the right shows the cluster volume of 15 example clusters as a function of time. Examples were recorded at glass age $t_{age}=10^5\tau_{LJ}$.}
\end{figure*}
To obtain a complementary picture of the \emph{spatially resolved} dynamics, we perform a spatial cluster analysis on the subset of hops in the same time window $[t_{age},t_{age}+t]$ that is used for the calculation of $N_{caged}$. We use a standard single-linkage cluster criterion, i.e., hops \textit{i} and \textit{j} are part of the same cluster, if the distance between the initial positions is below a threshold
\begin{equation*}
|\mathbf{r}_{init}^i-\mathbf{r}_{init}^j|<r_{cl}\; .
\end{equation*}
If a hop \textit{k} is already in a cluster with \textit{i}, then hops \textit{j} and \textit{k} will belong to the same cluster even if they don't fulfill the above criterion. As a threshold we use the spatial correlation range that is obtained from the density-density correlation $r_{cl}=r_{corr}=1.5\sigma$ (see previous section). As we increase the time window, we include more hops into the analysis and the clusters grow and merge. In Fig.~\ref{fig:growth} we illustrate the observed growth via snapshots of an example cluster at increasing time $t$. The cluster first consists only of a few hopping particles and successively grows into an extended structure. The grow process exhibits periods of near stagnation interrupted by large bursts which are due to the merging of simultaneously existing clusters. The plot in Fig.~\ref{fig:growth} illustrates this intermittent growth. It shows the cluster volume of 15 example clusters as a function of time. We define the volume of a cluster as the total correlated space of all hops that comprise the cluster
\begin{equation*}
V_{cl} = \cup_{i} V_{sp}(r_{init}^{i},r_{corr})\;.
\end{equation*}
Each hop contributes a spherical volume $V_{sp}$ with radius equal to the correlation range $r_{corr}=1.5\sigma$ centered around the initial position of the particle. In other words, with each hop we associate the volume of the entire cage and the cluster volume is the union of all cages that have rearranged. To calculate this joined volume, we use a voxel technique, i.e., we partition the simulation box into small cubes (voxels) and count the number of voxels with a center closer than $r_{corr}$ to any hop of a given cluster.

\begin{figure}[tb]
\includegraphics{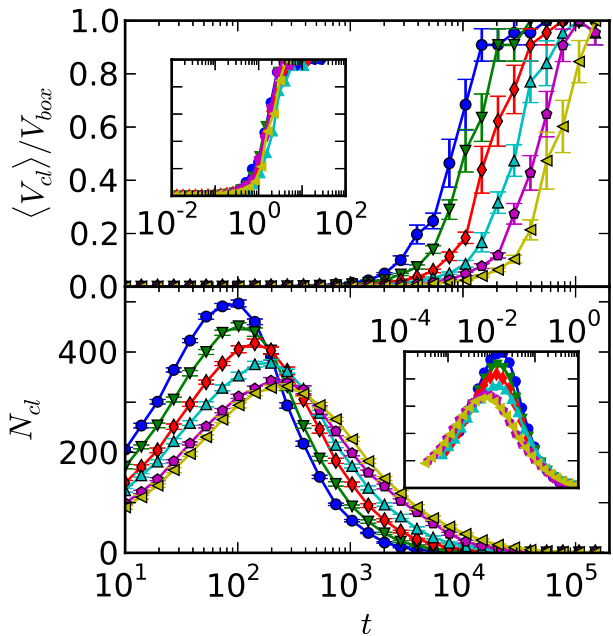}
\caption{\label{fig:clusterQuantities}(Color online) Mean cluster volume (top) given as a fraction of the total simulation box volume and the number of simultaneous clusters (bottom) as a function of time. Results for six ages are shown, see legend in Fig.~\ref{fig:chi4}. The insets show data collapse when time is rescaled by the time of the $\chi_4$ peak.}
\end{figure}
To explore the cluster configurations near the $\chi_{4}$ peak we calculate the mean volume and number of clusters as a function of time, and results for six ages are shown in Fig.~\ref{fig:clusterQuantities}. In the upper panel one can see that at long times we observe a single cluster that spans the simulation box, and its formation is shifted in time with increasing age. In the inset we show the same data with a time axis that is rescaled by the time of the $\chi_{4}$ peak, which collapses the data onto a single master curve. Note that the dominating cluster emerges just when the four-point susceptibility reaches its peak. The success of the scaling collapse indicates that aging merely delays, but does not otherwise alter the formation of this dominating cluster. The amount of clusters (bottom panel) peaks at much earlier times, and the rescaled data in the inset also shows an approximate collapse with age. Additionally, we find an age dependence of the peak height, showing that the maximal number of clusters decreases with increasing age.

\begin{figure}[tb]
\includegraphics{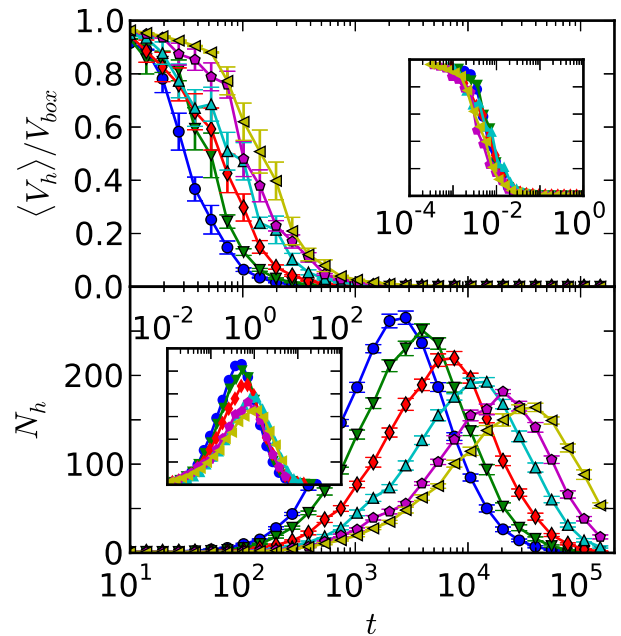}
\caption{\label{fig:holeQuantities}(Color online) Mean hole volume (top) given as a fraction of the total simulation box volume and the number of simultaneous holes (bottom) as a function of time. Results for six ages are shown, see legend in Fig.~\ref{fig:chi4}. The insets show data collapse when time is rescaled by the time of the $\chi_4$ peak.}
\end{figure}
To gain a more complete picture of the formed structures, we also measure the extent of the inactive regions in the glass where no hops are present. To quantify these "holes," we use the voxel partition (see above) and perform a nearest neighbor cluster analysis on the subset of voxels that does not lie inside the volume of any cluster. In Fig.~\ref{fig:holeQuantities} we show the mean volume of the holes (top panel) and the number of separate holes (bottom panel) as a function of time. We find that as the time window increases, the size of the holes shrinks, which is of course due to the growing hop clusters. Again, we observe a shift with glass age towards larger times, and indeed the inset in the top panel reveals that the break up of the single dominating inactive region happens just when the number of clusters is largest (both at $\sim 10^{-2}$ in rescaled time). In analogy, we find that the number of holes $N_{h}$ is maximal just when the mean cluster volume diverges and therefore when $\chi_{4}$ reaches its peak (see rescaled inset in the bottom panel).  The maximum in $N_{h}$ appears when the probability of \emph{closing} a hole by placing a new hop into the system becomes larger than the probability of \emph{splitting} a hole with a new hop. Therefore, the majority of the inactive regions have shrunk to a size on the order of a single cage at the time of the $\chi_4$-peak. The time of this crossover shows the same age-dependence as $\chi_4$, yet we also find that the maximal number of holes decreases with age. The scaling behavior of the hole volumes with age mirrors the behavior of the cluster volumes, and further supports the interpretation that the geometry of DH is unchanged by aging.

\begin{figure}[tb]
\includegraphics{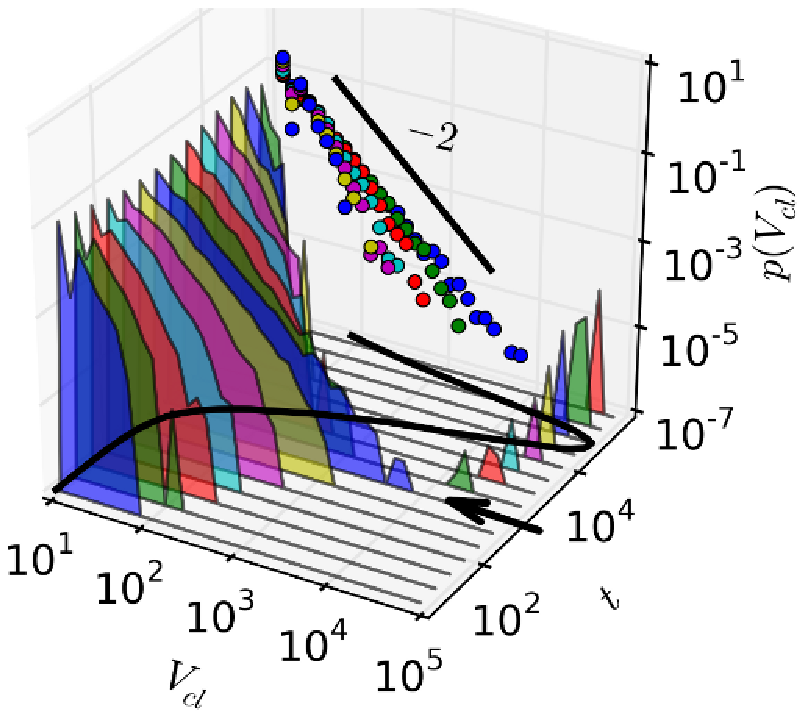}
\caption{\label{fig:volumeDist}(Color online) Collection of cluster volume distributions of a glass at age $t_{age}=10^5\tau_{LJ}$ measured at various times in double-log scale. Each distribution is plotted in the $V_{cl}$-$p(V_{cl})$ plane and placed along the $t$ axis according to the size of the time window used for the cluster analysis. The back wall shows an overlay of distributions at small times in a single plane. We include data at times $\lesssim 10^{3}\tau_{LJ}$ (up to and including the second blue distribution, see arrow) and the same colors as the separate distributions are used to indicate the origin of the data points. The solid line on the back wall indicates a power law with exponent $-2$. The black solid curve on the floor wall indicates $\chi_4$ as a function of time.}
\end{figure}
The mean cluster volume already suggests a single, dominating cluster in the system when $\chi_{4}$ is maximal. We gain further insight by directly studying the full distribution of cluster volumes. In Fig.~\ref{fig:volumeDist} we show its evolution for a single glass age, where the distributions for increasing time windows $[t_{age},t_{age}+t]$ are stacked along the $t$ axis. One can clearly see how the distribution lengthens over time until $\sim 10^{3}\tau_{LJ}$. At this point (blue to green, see arrow) the dominating cluster is formed, indicated by a detached peak at large volume and the successive shortening of the remaining distribution. The solid black curve on the floor wall indicates $\chi_{4}$ as a function of time, and as the large cluster grows, so does $\chi_{4}$. The peak is reached when the dominating cluster essentially covers the whole volume. Furthermore, we analyzed the form of the distributions, which prior to the emergence of the dominating cluster follow a power law. On the back wall of Fig.~\ref{fig:volumeDist} we show an overlay of these early distributions (see caption) in a single plane. The overlay shows that the cluster volume distribution lengthens until a power law with exponent of approximately $-2$ is reached.

\begin{figure}[tb]
\includegraphics{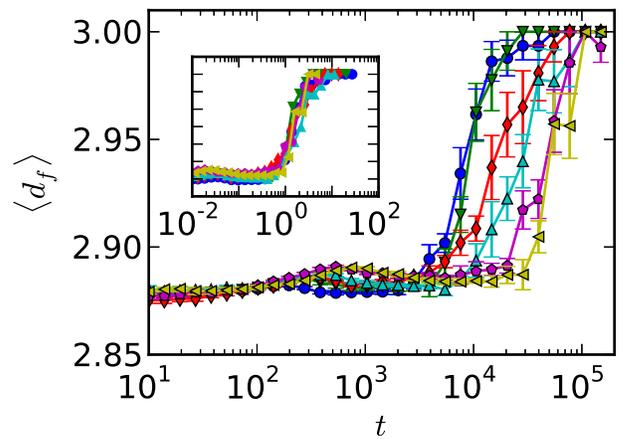}
\caption{\label{fig:fractalDim}(Color online) Main panel shows the mean fractal dimension of the hop clusters over time for six ages, see legend in Fig.~\ref{fig:chi4}. The inset shows data collapse when time is rescaled by the time of the $\chi_4$ peak.}
\end{figure}
From the snapshots in Fig.\ref{fig:growth}, one can see that the clusters are not compact, but have a complex geometry. In an earlier study, Parsaeian and Castillo~\cite{parsaeian_growth_2008} investigated DH in an aging binary LJ glass using four-point correlators. By assuming that the height of the $\chi_{4}$ peak was proportional to the correlated volume and using an identified correlation range $\xi$, they found a scaling of $\chi_{4}^{peak}\propto \xi^{b}$ with $b = 2.89\pm 0.03$. Our access to spatially resolved clusters allows us to directly calculate their fractal dimension and compare to this scaling result. We use the box counting method~\cite{falconer_fractal_2007}, in which one covers the cluster with successively smaller cubes, counting each time how many boxes are needed. Here, a cluster is represented by its correlated volume, i.e., the union of spheres with radius $1.5\sigma$ centered at each hop. The dimensionality of the cluster is then calculated via power-law fit
\begin{equation*}
N_{l} \propto l^{d_{f}}\; ,
\end{equation*}
where $l$ is the side length of the cube and $N_{l}$ the cube count. In Fig.~\ref{fig:fractalDim} we show the fractal dimension as a function of time for six ages. We find a mean fractal dimension that is stable around $2.88$ for short and intermediate times and increases to $3$ for long times. The inset shows that this increase happens as $\chi_{4}$ peaks and from the discussion of the volume distribution above, we know that the peak is accompanied by the emergence of a system spanning cluster. Since such a cluster will have the dimensionality of our simulation box, this increase to $d_{f}=3$ is not surprising and is clearly a finite size effect. Therefore, our value of $\langle d_{f} \rangle = 2.88$ agrees remarkably well with the above mentioned result that was solely based on four-point correlators.

\section{Conclusions}
We studied the microscopic dynamics of a quiescent 3D polymer glass in the aging regime, using a standard bead-spring model and molecular dynamics techniques. A refined version of a detection algorithm initially introduced by Candelier et al.~\cite{candelier_building_2009} was presented. It allowed us to efficiently measure local relaxation events defined as particle hops everywhere in the system and on-the-fly for the full duration of the simulation. An evaluation of the distribution of persistence times, hop distance, and hop autocorrelation at three temperatures showed good agreement with previous studies that used other methods~\cite{warren_atomistic_2009,vollmayr-lee_microscopic_2013}. Since our algorithm allows hop detection for the full system, we were able to present a spatio-temporal density-density correlation between relaxation events. A strong correlation was observed between near-simultaneous hops of neighboring particles, which indicates cooperative motion of groups of particles. We estimated correlation ranges and used these to analyze the size of the collaborative rearrangements as a function of temperature and age. We found distributions that first have an exponential shape and then transition over to a power-law tail that becomes flatter during aging. An increase in temperature broadened the power law, and this trend connects well to the power-law distributions seen by Candelier et al.~\cite{candelier_building_2009} in agitated granular media, where a very similar definition for the rearranging groups was used. An earlier study of a binary LJ-glass in the aging regime on the other hand showed power-law distributions~\cite{vollmayr-lee_self-organized_2006}, both for various temperatures and ages. We believe that this disagreement is due to the very different hop-time resolution, that was about three orders of magnitude smaller than what was used in this work.

In the second part of this study, we compared the standard $\chi_{4}$ measure of dynamical heterogeneity (DH) with a direct geometric analysis of hop clusters, which gives a spatially resolved picture to complement the bulk averaged $\chi_{4}$. Our results show that $\chi_{4}$ reaches its peak when a single dominating cluster is developed that extends throughout the system and is accompanied by mostly single-cage-sized pockets of inactive particles. We also observed a delayed cluster aggregation in older glasses that mirrored the shift of the $\chi_{4}$ peaks towards larger times with increasing age. Therefore, the geometric formation of DH is continuously slowed but otherwise unchanged by physical aging. We furthermore observed increasing $\chi_4$ peak heights, which indicate a growing dynamical correlation range during aging. Both the shift of $\chi_{4}$ with increasing age and the increasing $\chi_4$ peak height were also reported by Parsaeian and Castillo~\cite{parsaeian_growth_2008} in simulations of a binary LJ mixture. Recently, further evidence for growing dynamical correlations was obtained via experimental measurements of the nonlinear dielectric susceptibility in glycerol by Brun et al.~\cite{brun_evidence_2012}. Parsaeian and Castillo also identified a power-law scaling between an estimated growing correlation range and the $\chi_4$-peak height, which is connected to the total correlated volume. We showed that this scaling is in excellent agreement with the fractal dimension of the hop clusters. The mean cluster volume did not directly reveal the aging correlation range, as it is not proportional to $\chi_{4}$ in the range of its peak, yet a clear age dependence was observed for the maximal number of clusters and inactive regions (holes).

The shape of the evolving distribution of hop cluster volumes could help to understand the somewhat surprising success of mean-field models of aging. Despite the presence of heterogeneous dynamics, aging continuous time random walk descriptions~\cite{warren_atomistic_2009} based on the trap model of aging~\cite{monthus_models_1996} are very successful in capturing the evolution of mean squared displacements, dynamical structure factors and van Hove functions while entirely neglecting DH. Our measurements of the cluster volume distribution prior to the merging into a single dominating cluster (Fig.~\ref{fig:volumeDist}) showed a power-law form with exponent $\leq-2$. This observation indicates that fluctuations in the size of the DH are sufficiently small that average quantities such as mean cluster size do not behave anomalously.

The tools developed here for the detection of local relaxation events with high temporal as well as spatial resolution open up new possibilities in the study of glassy dynamics. Recently, Priezjev~\cite{priezjev_heterogeneous_2013} investigated the distribution of hops in a cyclically sheared glass with Candelier et al.'s algorithm and found intermittent bursts of hops driven by the strain amplitude as well as hop clusters with increasing size. An interesting avenue of future work would be to use our improved algorithm to extend the study of mechanically deformed glasses to larger systems and to directly measure the full distribution of cluster sizes.

\section*{Acknowledgments}
We thank J.-L. Barrat, R. Candelier, and O. Dauchot for helpful discussions. This work was supported by the Natural Sciences and Engineering Council of Canada (NSERC). Computing time was provided by WestGrid.

\bibliography{HopDistribution}

\end{document}